# Dzyaloshinskii-Moriya torque-driven resonance in antiferromagnetic $\alpha$-Fe$_2$O$_3$

*Qiyao Liu, Taeheon Kim, Kyusup Lee, Dongsheng Yang, Dushyant Kumar, Fanrui Hu and*

*Hyunsoo Yang[*]*

Department of Electrical and Computer Engineering, National University of Singapore

4 Engineering Drive 3, Singapore 117583

E-mail: eleyang@nus.edu.sg

We examine the high-frequency optical mode of $\alpha$-Fe$_2$O$_3$ and report that Dzyaloshinskii-Moriya (DM) interaction generates a new type of torque on the magnetic resonance. Using a continuous-wave terahertz interferometer, we measure the optical mode spectra, where the asymmetric absorption with a large amplitude and broad linewidth is observed near the magnetic transition point, Morin temperature ($T_\mathrm{M} \sim$ 254.3 K). Based on the spin wave model, the spectral anomaly is attributed to the DM interaction-induced torque, enabling to extract the strength of DM interaction field of 4 T. Our work opens a new avenue to characterize the spin resonance behaviors at an antiferromagnetic singular point for next-generation and high-frequency spin-based information technologies.

## 1. Introduction

With the emerging demand on high-frequency technologies, the spin resonance has been studied in antiferromagnets (AFMs), in the terahertz regime due to strong exchange interaction of sublattices.[1-2] A singular point with drastic magnetization change from collinear to spin-canting by applying a strong external magnetic field has attracted wide interest, such as in AFM Cr$_2$O$_3$ [3-5] and MnF$_2$ [6-7]. Hematite ($\alpha$-Fe$_2$O$_3$),[8] a non-collinear AFM under ambient conditions, undergoes the first-order phase transition at the Morin temperature of $T_\mathrm{M}$, leading to the temperature-dependent magnetic anisotropy. Above $T_\mathrm{M}$, the easy-axis phase changes to the easy-plane phase, hosting a canted spin structure with acoustic and optical resonant modes owing to Dzyaloshinskii-Moriya (DM) interaction.[9-10] Recent works on spin pumping in hematite have focused on the acoustic resonant mode (low-frequency < 38 GHz).[11-12] However, the optical resonant mode with hundreds of GHz is desirable for the next generation antiferromagnet-based spintronics. Despite the growing interest, studies of the optical mode have been limited in the excitation and detection of the resonance mode.[13] The underlying physics of the optical mode has not been fully resolved [14-15] and the enhanced resonant behaviors of hematite and orthoferrites near $T_M$ have been elusive.[16-18]

Even though multiple techniques such as neutron scattering,[19] terahertz transmission,[13, 20] and electron paramagnetic resonance[21] have been devoted to exploring the AFM resonance, little attention has been paid to dynamic details near $T_M$. It is attributed to severe challenges in measuring their linewidth induced by a low magnetic damping.[22-23] In this work, we implement a novel interferometric continuous wave-terahertz (CW-THz) spectroscopy methodology to measure the detail dynamics of a single crystal $\alpha$-Fe$_2$O$_3$. The phase-detection method based on an interferometer offers a high sensitivity with a better spectral resolution, with which the laser interferometer gravitational-wave observatory (LIGO) successfully detected cosmic gravitational waves.[24] Asymmetric spectra with a larger absorption amplitude and a broad linewidth near the $T_M$ are observed, which differs from that of the acoustic resonant mode.[11-12] Using the antiferromagnetic spin-wave model, we derive the theoretical response function driven by the THz magnetic field and DM interaction. The THz magnetic field induces a symmetric resonance spectrum, and the amplitude of THz magnetic field-induced resonance is independent of the resonance frequency. In contrast, the DM interaction induces an anti-symmetric resonance spectrum, and the amplitude of DM interaction-induced resonance is inversely proportional to the resonance frequency, and thus is responsible for the enhanced amplitude and the broadened linewidth near $T_M$. The above two different torque components enable to extract the strength of DM interaction. Our work marks a comprehensive exploration of the singular sub-THz response of hematite in the vicinity of Morin transition.

## 2. Results and Discussion

We employ a 1 mm thick $\alpha$-Fe$_2$O$_3$ crystal with [01$\bar{1}$2] cut (R-cut) to easily find the phase transition point and explore the excitation mechanism in both spin configurations. Figure 1a shows the hexagonal corundum crystal structure, opposite spin layers stacking alternatively along the direction perpendicular to the (0001) basal plane. By measuring the in-plane saturation magnetization, $T_M$ is found to be ~254.3 K, as shown in Figure 1b. The spin configurations accordingly change across $T_M$. The Néel angle $\theta_0$ changes from 0° ($T < T_M$) to 90° ($T > T_M$) at equilibrium, where $\theta_0$ is the polar angle of the Néel vector $\mathbf{l} = (\mathbf{M}_1 - \mathbf{M}_2)/2$ from the $z$ axis, and $\mathbf{M}_1$ and $\mathbf{M}_2$ represent the sublattice magnetizations as depicted in Figure 1b. The net magnetization $\mathbf{m} = (\mathbf{M}_1 + \mathbf{M}_2)/2$ is negligibly small for $T < T_M$, which increases substantially for $T > T_M$. The in-plane anisotropy of the canted AFM at room temperature is determined by measuring the magnetic hysteresis loops along with different azimuthal angles. The extracted coercive field ($H_c$) in Figure 1c shows an uniaxial in-plane anisotropy (see more information in Section 1, Supporting Information).

Similar to the ferromagnetic resonance phenomenon, the antiferromagnetic resonance occurs through the phase matching between the oscillating magnetic field of THz wave and the collective precession of spins. Here, we adopt the CW THz interferometer as shown in Figure 2a in AFM resonance detection for the first time (see *Experimental Section* for details). The captured THz intensity $I_{THz}$, which is dependent on the amplitude of the THz electric field $A_{THz}$ and the phase difference $\Delta\varphi$ between the two converged beams from two arms, can be described as

$$I_{THz} \propto \frac{1}{2} A_{THz}^2 [\cos(\Delta\varphi) + 1] = \frac{1}{2} A_{THz}^2 [\cos(4\pi \Delta L \nu / c) + 1], \tag{1}$$

where $\nu$ is the THz frequency, $c$ is the speed of light, and $\Delta L = L_1 - [L_2 + (n_0 - 1)d]$ represents the optical length difference between the two arms ($L_1$ and $L_2$) of the interferometer, including the optical path length ($n_0 d$) of a sample (Section 2, Supporting Information).

Through changing $\Delta L$, the THz intensity in Eq. (1) is modulated with the relative phase difference for a given frequency. As a result, the detector records series of interference at different frequencies with the delay stage scanning, each carrying a specific phase. Figure 2b shows the captured data and sinusoidal fits at 200.0, 200.2, and 200.4 GHz, where the period $\Lambda = c/(2\nu)$ agrees with Eq. (1). The phase $\varphi$ can be extracted at each given frequency from the fitting results. Further phase correction is carried out since the extracted raw phase data are wrapped within $2\pi$ range (Section 3, Supporting Information).[25-26] The absorption can be obtained from the phase by the Kramers-Kronig (K-K) relationship (Section 4, Supporting Information).[27-28]

First, we perform the phase detection in $\alpha$-Fe$_2$O$_3$ from 296 to 350 K as shown in Figure 2c. Since the refractive index $n_0$ of $\alpha$-Fe$_2$O$_3$ in the THz range is almost independent of frequency ($n_0 \sim 5.5$),[29] the phase $\varphi(\nu) \sim (n_0 - 1)\nu$ is linear with the frequency. Three obvious kinks happen at ~178, 223, and 268 GHz (indicated by grey shadows), which are caused by Fabry-Pérot resonance absorption at the high-resistive silicon beam splitter (BS). The kinks do not shift as sample temperature changes, and their period (~45 GHz) is determined by the effective optical path of BS. Meanwhile, there is a phase anomaly that shifts systematically with temperature labeled with red arrows in Figure 2c, whose behavior is consistent with the high-frequency resonance mode in hematite in the easy-plane phase for $T > T_M$. We then perform measurements in a broader range of temperatures down to 78 K, and the phase anomalies are captured accordingly. Figure 2d shows the extracted absorptions in various temperatures, where the scattered squares are data points, and the line curves are Lorentzian fits.

The temperature dependence of the resonance frequency is summarized in **Figure 3**a. As temperature rises, the resonance frequency of 209 GHz at 78 K decreases down to 78 GHz near $T_M$ and then increases for $T > T_M$. Near the Morin temperature, the thermodynamic properties of displacive phase transitions are related to instabilities arising in collective modes of the system.[30] The observed temperature dependence of the resonance frequency agrees well with the high-frequency magnetic resonant mode in hematite,[13, 20] which is further supported by the angle-dependent THz excitation measurements and spin wave theory (Section 5 and Section 6, Supporting Information).

Figure 3b shows the temperature dependence of linewidth, which is defined as the full width at half-maximum of a Lorentzian fit for the absorption. The narrow linewidths < 3 GHz over the entire temperature range imply the low magnetic damping in both easy-plane and easy-axis phases, thus reflecting a long magnon decay length in hematite. At $T > 275$ K, the broader linewidth with elevating temperature is due to a thermal effect that increases the scattering cross-section of the two-magnon process.[31-32] For $T < 180$ K, the linewidth increases when the temperature decreases, which is attributed to the presence of a small admixture of $Fe^{2+}$ in this natural crystal, inducing freezing of the electronic $Fe^{2+}$-$Fe^{3+}$ transitions.[16] When approaching $T_M$, the linewidth shows a peak, which can be explained by stronger dissipation processes due to the minimum anisotropy at the Morin transition.[21]

The $Q$ factor, defined as the resonance frequency divided by the linewidth, is evaluated in Figure 3c. The maximum value of 570 is obtained at 180 K with a resonance frequency of 182.5 GHz, in which the linewidth shows the minimum value of 0.32 GHz. From the viewpoint of application in spin-based informatics, although the $Q$ factors of magnetic materials have reached several thousands in microwave frequencies, the development and exploration in THz range lags.[12] In previous studies, the $Q$ factors in most orthoferrites were restricted to several tens.[33] Exceptionally, the $Q$ factor in $YFeO_3$ has been reported up to 332,[34] and that of an $ErFeO_3$ crystal has been recently found to exceed 1000.[33] However, as the most prevalent and thermodynamically stable oxide under ambient conditions, little attention has been paid to the $Q$ factor of spin precession in $\alpha$-$Fe_2O_3$ in a wide temperature range. It is worth noting that recent studies have realized tailoring the Morin temperature of hematite to room temperature by a doping technique using $Rh^{3+}$ and $Ru^{3+}$.[35-36] Based on this development and our findings, $\alpha$-$Fe_2O_3$ is a promising candidate to provide a low phase noise in THz spin oscillators and filters.

In addition, we find that the resonance amplitude in Figure 3d shows a distinct peak near $T_M$. While a similar amplitude behavior was reported in another canted AFM, $TmFeO_3$ crystal, it is vaguely attributed to the inverse correlation between the AFM resonance amplitude frequency.[37] Such singular behaviors in the amplitude and linewidth near the Morin transition have been little explored based on spin dynamics nor understood.

In order to understand the enhancement of the amplitude and linewidth near the transition temperature, we use a macroscopic uniform $k = 0$ spin-wave model (Section 7, Supporting Information) with ansatz $\mathbf{l} = (0, l_y, l_z) = (0, \sin\theta, \cos\theta)$,

$$\ddot{\theta} + 2\omega_{Ex}\beta\dot{\theta} + \omega_{LR}^2 \sin(2\theta)/2 = F_{THz} + F_{DM} \qquad (2)$$

where $\theta$ is the Néel angle, $\omega_{Ex}$ is the angular frequency converted from the exchange field ($H_{Ex}$) by multiplying the gyromagnetic ratio $\gamma$ ($\omega_{Ex} = \gamma H_{Ex}$), $\beta$ is the phenomenological damping constant and $\omega_{LR}$ is the linear resonance angular frequency consisting of the exchange field, DM field ($H_{DM}$), and two temperature-dependent uniaxial anisotropy fields along the $z$ axis.

The generalized driving forces consist of two terms. The first one is the conventional driving force exerted by the THz magnetic field of light, $F_{THz} = \gamma \dot{H}_{THz}$ where $H_{THz} = H_{ac}\sin(\omega t)$ and $F_{THz} = \gamma \frac{d(H_{ac}\sin(\omega t))}{dt} = \gamma\omega H_{ac}\cos(\omega t)$. Therefore, the strength of $F_{THz}$ is proportional to $\omega$. The second one is the DM interaction-induced driving force $F_{DM}$. It is simplified as $F_{DM} = -\gamma^2 H_{DM} H_{THz} \cos(\theta) = -\gamma^2 H_{DM} H_{ac} \sin(\omega t)\cos(\theta)$. The activation of $F_{DM}$ subsequently follows the excitation by $H_{THz}$, during which the precession caused by $H_{THz}$ induces a perpendicular component to the DM vector, and thereby, $F_{DM}$ induces further oscillation. Therefore, the torque strength of $F_{DM}$ is independent of $\omega$, maximum for the collinear AFM state ($\theta_0 = 0°$), and zero for the canted AFM state ($\theta_0 = 90°$).

Figure 4a shows the response function $R(v) = \frac{1}{\tau}\int_0^\tau \Delta\theta(t,v)dt$, numerically calculated using Eq. (2), where $\tau = 2\pi/\omega = 1/v$ and $\Delta\theta(t,v)$ is the precessional angle of Néel vector. When a weak magnetic field of THz wave excites α-Fe$_2$O$_3$, the time-averaged amplitude of the Néel vector is $R(v)$, which is regarded as absorption. At $T < T_M$ and $T > T_M$, the resonance amplitude, defined as the maximum amplitude of the response function envelope, is small, while the resonance amplitude becomes large near $T_M$. In simulations, the $T_M$ of 254.3 K is estimated as a crossing point between two branches of frequency shift with temperature [30] (Section 6, Supporting Information). Since the resonance amplitude depends on the potential curvature or resonance frequency $\omega_{LR}(\theta) \equiv \frac{\partial^2 V(\theta)}{\partial \theta^2}$, [38] we plot the magnetic potential $V(\theta) \equiv -\int \omega_{LR}^2 \sin(2\theta)/2 d\theta$ at 78, 254.3 ($T_M$), and 330 K, with spin trajectories of the sublattices (**M**$_1$ and **M**$_2$) and the Néel vector (**l**) as shown in insets of Figure 4a. At $T = 78$ K $< T_M$, the **l** oscillates around the equilibrium Néel angle $\theta_0 = 0°$ as shown in the inset of the bottom panel in Figure 4a, where $\left.\frac{\partial V(\theta)}{\partial \theta}\right|_{\theta=\theta_0} = 0$. When $T = 330$ K $> T_M$, $\theta_0$ reorients from 0° to 90°, and **l** oscillates around $\theta_0 = 90°$ as shown in the inset of the top panel in Figure 4a. Due to the high potential curvature, the oscillation is strongly suppressed for both $T < T_M$ and $T > T_M$ cases. On the other hand, a strong resonant oscillation emerges when the potential profile shows a low curvature near $T_M$ as shown in the middle panel in Figure 4a.

To analyze the detailed contribution of $F_{DM}$ and $F_{THz}$ to $R(v)$, we derive the response function with spin wave ansatz $\mathbf{l} = (A_\varphi e^{i\omega t}, A_\theta e^{i\omega t}, 1)$ at $T < T_M$ or $\mathbf{l} = (A_\varphi e^{i\omega t}, 1, A_\theta e^{i\omega t})$ at $T > T_M$. Thus, for $T < T_M$

$$\begin{pmatrix} H_{ac}(i\omega + \omega_{DM}) \\ 0 \end{pmatrix} = \begin{pmatrix} \omega^2 - \omega_{LR}^2 - 2i\beta\omega\omega_{Ex} & 0 \\ 0 & \omega^2 - i\omega\omega_{DM} - 2\omega_{Ex}(\omega_{K2} + \omega_{K4}) - 2i\beta\omega\omega_{Ex} \end{pmatrix} \begin{pmatrix} A_\theta \\ A_\varphi \end{pmatrix} \quad (3)$$

By solving Eq. (3), we obtain

$$\begin{pmatrix} A_\theta \\ A_\varphi \end{pmatrix} = \begin{pmatrix} \dfrac{4\omega_{Ex}^2}{\omega^2 - \omega_{LR}^2 - 2i\beta\omega\omega_{Ex}} & 0 \\ 0 & \dfrac{4\omega_{Ex}^2}{\omega^2 - i\omega\omega_{DM} - 2\omega_{Ex}(\omega_{K2} + \omega_{K4}) - 2i\beta\omega\omega_{Ex}} \end{pmatrix} \begin{pmatrix} H_{ac}(i\omega + \omega_{DM}) \\ 0 \end{pmatrix} \quad (4).$$

The response function $R(v)$ is obtained by taking the time average to the real part of $A_\theta$ in Eq. (4). It decomposes into two components

$$R(v) = \frac{1}{\tau} \int_0^\tau \mathrm{Re}[A_\theta(v)] dt$$

$$= S \frac{2 H_{ac} v^2 \beta \gamma H_{Ex}}{(2\pi)^2 (v^2 - v_{LR}^2)^2 + (2\beta v \gamma H_{Ex})^2} + A \frac{H_{ac}(v^2 - v_{LR}^2) \gamma H_{DM}}{(2\pi)^2 (v^2 - v_{LR}^2)^2 + (2\beta v \gamma H_{Ex})^2} \quad (5)$$

where $v = \omega/(2\pi)$ and $\tau = 1/v$. The symmetric $S$ (the antisymmetric $A$) component is induced by the $F_{THz}$ ($F_{DM}$). For $T > T_M$, we obtain

$$R(v) = S \frac{2 H_{ac} v^2 \beta \gamma H_{Ex}}{(2\pi)^2 (v^2 - v_{LR}^2)^2 + (2\beta v \gamma H_{Ex})^2}, \quad (6)$$

Figure 4b shows $R(v)$ consisting of the $S$ and $A$ components at 78, 254.3, and 330 K. Note that the peak amplitude of the $S$ component at a frequency $v = v_{LR}$, defined as $R(v_{LR}) = S \dfrac{H_{ac}}{2\beta\gamma H_{Ex}}$, is constant and independent of $v_{LR}$ for all ranges of temperature. Although the $F_{THz}$ strength is stronger (weaker) in the system with a high (low) resonance frequency, the high (low) potential curvature suppresses (enhances) oscillation, and thereby the peak amplitude of $S$ component remains constant. We find that, on the other hand, the $A$ component has a large peak amplitude near $T_M$. For example, the peak amplitude of the $A$ component, defined near $v_{LR}$ as $|R(v_{LR} \pm C)| \sim \left| A \dfrac{H_{ac} \gamma H_{DM} C}{(4\pi v_{LR})(\beta\gamma H_{Ex})^2} \right|$ where $|v - v_{LR}| \sim C$ ($\ll 1$), is inversely proportional to $v_{LR}$. It indicates that the resonance enhancement near $v_{LR}$ originated from the contribution of DM interaction.

Notably, it is possible to extract the strength of DM interaction using the antisymmetric component of the response function. The spectral deconvolution of the experimental absorption data at $T_M$ is shown in Figure 4c, where the $A$ component induces an antisymmetric spectrum around $\nu_{LR}$ = 78.5 GHz. To extract the DM interaction energy, we exploit a simple relation between the spin canting angle $\eta$ of $Fe^{3+}$ ions and $M_s$: $\tan^{-1}\eta \sim H_{DM}/(2H_{Ex}) \sim M_s/M_0 \sim 0.0012$ where $M_0$ is the magnetic moment of $Fe^{3+}$ ions per unit mass. From the fitting parameters of $M_s$ = 0.4 emu/g, $M_0$ = 345 emu/g, $\nu_{LR}$ = 78.45 GHz, and $\beta = 8 \pm 1.4 \times 10^{-6}$, $H_{DM}$ and $H_{Ex}$ are estimated as $4 \pm 0.43$ T and ~ 833 T, respectively, which is in good agreement with the previous reports.[11-12, 16, 39]

Figure 4d shows the resonance amplitude without and with $F_{DM}$, confirming that $F_{DM}$ is responsible for the resonance enhancement as $T$ approaches $T_M$, whose driving force $F_{DM}$ intensifies the oscillation of **l** greatly due to a low potential curvature around $T_M$. Since $F_{DM}$ is maximum (zero) when $T \sim T_M$ ($T > T_M$) and $\theta_0 = 0°$ ($\theta_0 = 90°$), the amplitude response shows an asymmetric temperature characteristic, as shown in Figure 4d.

In addition, the different influences of $F_{DM}$ in two different magnetic phases are analyzed with a simplified model (Section 8, Supporting Information). When $T < T_M$, both **l** (as well as $\mathbf{M}_1$ and $\mathbf{M}_2$) and the DM vector **D** are parallel to the $z$ axis without $H_{THz}$, therefore $\mathbf{H}_{DM}$ is zero ($\mathbf{H}_{DM} = \mathbf{D} \times (-1)^{i+1}\mathbf{M}_i, (i=1,2)$). Once **l** is tilted slightly away from the $z$ axis by $H_{THz}$ applied along the $x$ axis, $F_{DM}$ is non-zero and induces a finite $\Delta\theta$ around $\theta_0$. In other words, $\mathbf{H}_{DM}$ exerts antisymmetric exchange torques on the $x$ components of sublattices induced by $H_{THz}$ ($\frac{d\mathbf{M}_i}{dt} \sim \mathbf{M}_i \times \mathbf{H}_{DM}, (i=1,2)$). When $T > T_M$ without $H_{THz}$, $\mathbf{M}_1$ and $\mathbf{M}_2$ are aligned in the $xy$ plane. $F_{DM}$ exerts on both $\mathbf{M}_1$ and $\mathbf{M}_2$, increasing **m**, but with a negligible $\theta$ change. Such an anisotropic feature with respect to temperature is qualitatively captured in the experimental data of Figure 3d. In addition, the significant oscillation behavior by $F_{DM}$ near $T_M$ corresponds to the broad linewidth in the absorption spectrum in Figure 4e. Comparing theory to the experimental results in Figure 3, we notice that the enhancements of both the amplitude and linewidth in the experiment occur in a much wider temperature range than that of simulations, and the dramatic decrease expected near $T_M$ in the simulation is moderated in the experimental observations. Such a phenomenon is understandable, considering the high sensitivity of spin reorientation to the derivations in crystalline perfection. Besides, the factor of thermal effect and dissipation process is difficult to be taken into account in equations. Therefore, the linewidth increase in $T > T_M$ range is not reproduced in simulations, which can be attributed to thermal smearing effects and sample inhomogeneities.[37, 40] We anticipate even stronger resonance to be observed in other canted AFMs, such as rare-earth orthoferrites, in which the DM strength is an order of magnitude stronger than hematite.[34]

## 3. Conclusion

In summary, we have investigated the high-frequency resonance mode in hematite $\alpha$-Fe$_2$O$_3$. The antisymmetric response function is attributed to DM interaction, which determines the strength and direction of the DM vector. While the magnetic field-induced driving force results in a constant resonance amplitude regardless of resonance frequency $\nu_{LR}$, the DM interaction-induced driving force ($F_{DM}$) amplifies the resonance amplitude, which is inversely proportional to the $\nu_{LR}$. Such dependence can be exploited in designing energy-efficient spin wave resonators with a high amplitude based on AFM for magnonics and microwave applications. We have also shown that the linewidth is enhanced by $F_{DM}$ similar to the amplitude. Finally, we estimate the strength of DM interaction from the spectrum analysis. Our findings shed light on the intriguing physics of the optical mode and are promising for understanding the influence of DM interaction on the magnetic soliton dynamics in canted AFMs.

## 4. Experimental Section

*Experimental details:* We use the transmission scheme in the CW interferometer to increase the response of incident THz wave on resonance. The data acquisition scheme in this work is realized by Michelson interferometry with phase detection. It is robust against the surrounding noises, and no additional limitation on spectral resolution is introduced by the delay stage. The hematite crystal was purchased from MaTecK GmbH, Germany. The CW terahertz beam was radiated from the photomixer (TeraScan 1550) by optical heterodyning, with the frequency exactly at the difference frequency of two tunable infrared lasers and spectral resolution < 10 MHz.[41] The divergent THz wave was divided into two arms by a 45° positioned 1 mm-thick high-resistive silicon plate, which functioned as a beam splitter (BS); one branch was directly reflected by a gold-coated reflector ($R_1$), as a reference, and the other one was closely focused onto the sample by a parabolic mirror with 2 inches focus length. The transmitted beam was then reflected by another plane mirror ($R_2$), whose position could be adjusted by a precise delay stage. Then the reflected beams of two arms converged at the beam splitter, interfered with each other, and were finally detected by a THz detector (a Schottky receiver with high responsivity in the range of 50-1500 GHz). During the whole experiment, the THz beam path in the interferometer was purged with dry air to minimize the absorption of THz radiation by atmospheric water vapor, and the humidity was kept below 10%.

*Statistical Analysis:* The dielectric tensor is modulated by the magnetic field of the incident wave due to the magneto-optical interaction, which corresponds to the variation of the complex refractive index $\boldsymbol{n} = \sqrt{\varepsilon\mu} = n_0 + i\kappa$ for absorption media.[42] Here $\varepsilon$ and $\mu$ are the complex dielectric constant and magnetic permeability of the substance, respectively; the real part $n_0$ is the refractive index indicating the phase velocity, while the imaginary part $\kappa$ denotes the extinction coefficient reflecting the absorption. The real and imaginary parts are connected by the K-K relationship, which allows deducing the absorption from phase information. The data processing method is as follows: First, the phase difference $\Delta\psi$ introduced by the spin resonance at temperature $T_1$ should be obtained. One can use the extracted phase at $T_1$ to deduct the reference phase at another temperature $T_2$. Second, we directly proceed with K-K transformation to $\Delta\psi$, which is proportional to $n_0$. Therefore, the absorption feature ($\sim \kappa$) can be achieved. Third, we use Lorentzian fitting to extract the linewidth and relative amplitude of absorption.

**Acknowledgements**

Q. Liu and T. Kim contributed equally to this work. The research is supported by National Research Foundation (NRF) Singapore Investigatorship (NRFI06-2020-0015).

**References**

[1] V. Baltz, A. Manchon, M. Tsoi, T. Moriyama, T. Ono, Y. Tserkovnyak, *Rev. Mod. Phys.* **2018**, 90, 015005.
[2] L. N. Kapoor, S. Mandal, P. C. Adak, M. Patankar, S. Manni, A. Thamizhavel, M. M. Deshmukh, *Adv. Mater.* **2021**, 33, e2005105.
[3] J. Li, C. B. Wilson, R. Cheng, M. Lohmann, M. Kavand, W. Yuan, M. Aldosary, N. Agladze, P. Wei, M. S. Sherwin, J. Shi, *Nature* **2020**, 578, 70.
[4] S. Seki, T. Ideue, M. Kubota, Y. Kozuka, R. Takagi, M. Nakamura, Y. Kaneko, M. Kawasaki, Y. Tokura, *Phys. Rev. Lett.* **2015**, 115, 266601.
[5] W. Yuan, Q. Zhu, T. Su, Y. Yao, W. Xing, Y. Chen, Y. Ma, X. Lin, J. Shi, R. Shindou, X. C. Xie, W. Han, *Sci. Adv.* **2018**, 4, eaat1098.
[6] P. Vaidya, S. A. Morley, J. van Tol, Y. Liu, R. Cheng, A. Brataas, D. Lederman, E. del Barco, *Science* **2020**, 368, 160.
[7] S. M. Wu, W. Zhang, A. Kc, P. Borisov, J. E. Pearson, J. S. Jiang, D. Lederman, A. Hoffmann, A. Bhattacharya, *Phys. Rev. Lett.* **2016**, 116, 097204.
[8] A. H. Morrish, *Canted antiferromagnetism: hematite*, World Scientific, **1994**.
[9] I. Dzyaloshinsky, *J. Phys. Chem. Solids* **1958**, 4, 241.
[10] X. Liu, Q. Feng, D. Zhang, Y. Deng, S. Dong, E. Zhang, W. Li, Q. Lu, K. Chang, K. Wang, *Adv. Mater.* **2023**, 2211634.
[11] H. Wang, Y. Xiao, M. Guo, E. Lee-Wong, G. Q. Yan, R. Cheng, C. R. Du, *Phys. Rev. Lett.* **2021**, 127, 117202.
[12] I. Boventer, H. T. Simensen, A. Anane, M. Kläui, A. Brataas, R. Lebrun, *Phys. Rev. Lett.* **2021**, 126, 187201.
[13] K. Grishunin, E. A. Mashkovich, A. V. Kimel, A. M. Balbashov, A. K. Zvezdin, *Phys. Rev. B* **2021**, 104.
[14] M. Białek, J. Zhang, H. Yu, J.-P. Ansermet, *Appl. Phys. Lett.* **2022**, 121, 032401.
[15] M. Białek, J. Zhang, H. Yu, J.-P. Ansermet, *Phys. Rev. Appl.* **2021**, 15, 044018.
[16] K. S. Aieksandrov, Bezrnaternykh, L. N., Kozlov, G. V., Lebedev, S. P., Mukhin, A. A. and Prokhorov, A. S., *Sov. Phys. JETP* **1987**, 65, 591.
[17] A. V. Kimel, B. A. Ivanov, R. V. Pisarev, P. A. Usachev, A. Kirilyuk, T. Rasing, *Nat. Phys.* **2009**, 5, 727.
[18] J. Tang, Y. Ke, W. He, X. Zhang, W. Zhang, N. Li, Y. Zhang, Y. Li, Z. Cheng, *Adv. Mater.* **2018**, 30, e1706439.
[19] E. J. Samuelsen, G. Shirane, *Phys. Stat. Sol. B* **1970**, 42, 241.
[20] S. G. Chou, P. E. Stutzman, S. Wang, E. J. Garboczi, W. F. Egelhoff, D. F. Plusquellic, *J. Phys. Chem. C* **2012**, 116, 16161.
[21] R. Lebrun, A. Ross, O. Gomonay, V. Baltz, U. Ebels, A. L. Barra, A. Qaiumzadeh, A. Brataas, J. Sinova, M. Kläui, *Nat. Commun.* **2020**, 11, 6332.
[22] A. Little, L. Wu, P. Lampen-Kelley, A. Banerjee, S. Patankar, D. Rees, C. A. Bridges, J. Q. Yan, D. Mandrus, S. E. Nagler, J. Orenstein, *Phys. Rev. Lett.* **2017**, 119, 227201.
[23] J. Kiessling, R. Sowade, I. C. Mayorga, K. Buse, I. Breunig, *Rev. Sci. Instrum.* **2011**, 82, 026108.
[24] B. P. Abbott, et.al., *Phys Rev Lett* **2016**, 116, 131103.
[25] J. Gass, A. Dakoff, M. K. Kim, *Opt. Lett.* **2003**, 28, 1141.
[26] X. Wang, L. Hou, Y. Zhang, *Applied Physics* **2010**, 49, 5095.
[27] K. Peiponen, E. M. Vartiainen, *Phys. Rev. B* **1991**, 44, 8301.
[28] K. A. Whittaker, J. Keaveney, I. G. Hughes, C. S. Adams, *Phys. Rev. A* **2015**, 91, 032513.
[29] Z. Xu, D. Ye, J. Chen, H. Zhou, *Coatings* **2020**, 10, 805.
[30] L. M. Levinson, M. Luban, S. Shtrikman, *Phys. Rev.* **1969**, 187, 715.
[31] T. Moriyama, K. Hayashi, K. Yamada, M. Shima, Y. Ohya, T. Ono, *Phys. Rev. Mater.*

**2019**, 3, 051402.

[32] M. G. Cottam, *J. Phys. C Solid State Phys.* **1972**, 5, 1461.

[33] H. Watanabe, T. Kurihara, T. Kato, K. Yamaguchi, T. Suemoto, *Appl. Phys. Lett.* **2017**, 111, 092401.

[34] G. V. Kozlov, S. P. Lebedev, A. A. Mukhin, A. S. Prokhorov, I. V. Fedorov, A. M. Balbashov, I. Y. Parsegov, *IEEE Trans. Magn.* **1993**, 29, 3443.

[35] H. Jani, J. Linghu, S. Hooda, R. V. Chopdekar, C. Li, G. J. Omar, S. Prakash, Y. Du, P. Yang, A. Banas, K. Banas, S. Ghosh, S. Ojha, G. R. Umapathy, D. Kanjilal, A. Ariando, S. J. Pennycook, E. Arenholz, P. G. Radaelli, J. M. D. Coey, Y. P. Feng, T. Venkatesan, *Nat. Commun.* **2021**, 12, 1668.

[36] K. Hayashi, K. Yamada, M. Shima, Y. Ohya, T. Ono, T. Moriyama, *Appl. Phys. Lett.* **2021**, 119, 032408.

[37] A. V. Kimel, A. Kirilyuk, A. Tsvetkov, R. V. Pisarev, T. Rasing, *Nature* **2004**, 429, 850.

[38] T. H. Kim, P. Grünberg, S. H. Han, B. K. Cho, *Phys. Rev. B* **2018**, 97, 184427.

[39] R. Lebrun, A. Ross, S. A. Bender, A. Qaiumzadeh, L. Baldrati, J. Cramer, A. Brataas, R. A. Duine, M. Kläui, *Nature* **2018**, 561, 222.

[40] C. D. Stanciu, A. V. Kimel, F. Hansteen, A. Tsukamoto, A. Itoh, A. Kirilyuk, T. Rasing, *Phys. Rev. B* **2006**, 73, 220402.

[41] A. J. Deninger, T. Gobel, D. Schonherr, T. Kinder, A. Roggenbuck, M. Koberle, F. Lison, T. Muller-Wirts, P. Meissner, *Rev. Sci. Instrum.* **2008**, 79, 044702.

[42] I. S. Akhmadullin, V. A. Golenishchev-Kutuzov, S. A. Migachev, M. F. Sadykov, *Phys. Solid State* **2002**, 44, 333.

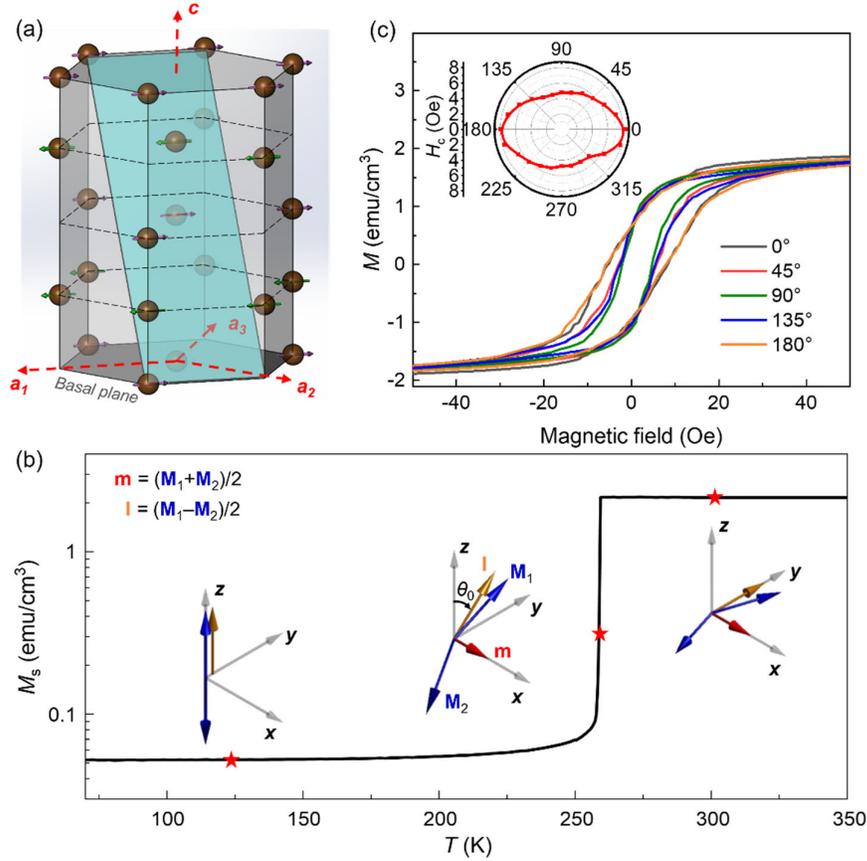

**Figure 1.** Crystalline configuration and magnetic properties in α-Fe$_2$O$_3$. a) Schematic of the α-Fe$_2$O$_3$ hexagonal lattice (above the Morin temperature) with Fe$^{3+}$ ions on basal plane and AFM coupling between adjacent planes (oxygen atoms are not shown), where the light blue plane represents the sample cutting plane of $(01\bar{1}2)$. b) In-plane saturation magnetization values ($M_s$) as a function of temperature. The insets show the corresponding spin configurations at three different temperatures (marked with stars), where the *xy* plane coincides with the crystal basal plane. The blue arrows denote two spin moments (**M**$_1$ and **M**$_2$), the red ones represent the net magnetization **m** = (**M**$_1$+**M**$_2$)/2, and the orange ones denote the Néel vector **l** = (**M**$_1$−**M**$_2$)/2. c) Magnetic hysteresis loops at room temperature. The in-plane anisotropy reflected by the coercive field ($H_c$) dependent on the sample azimuthal angles is shown in the inset.

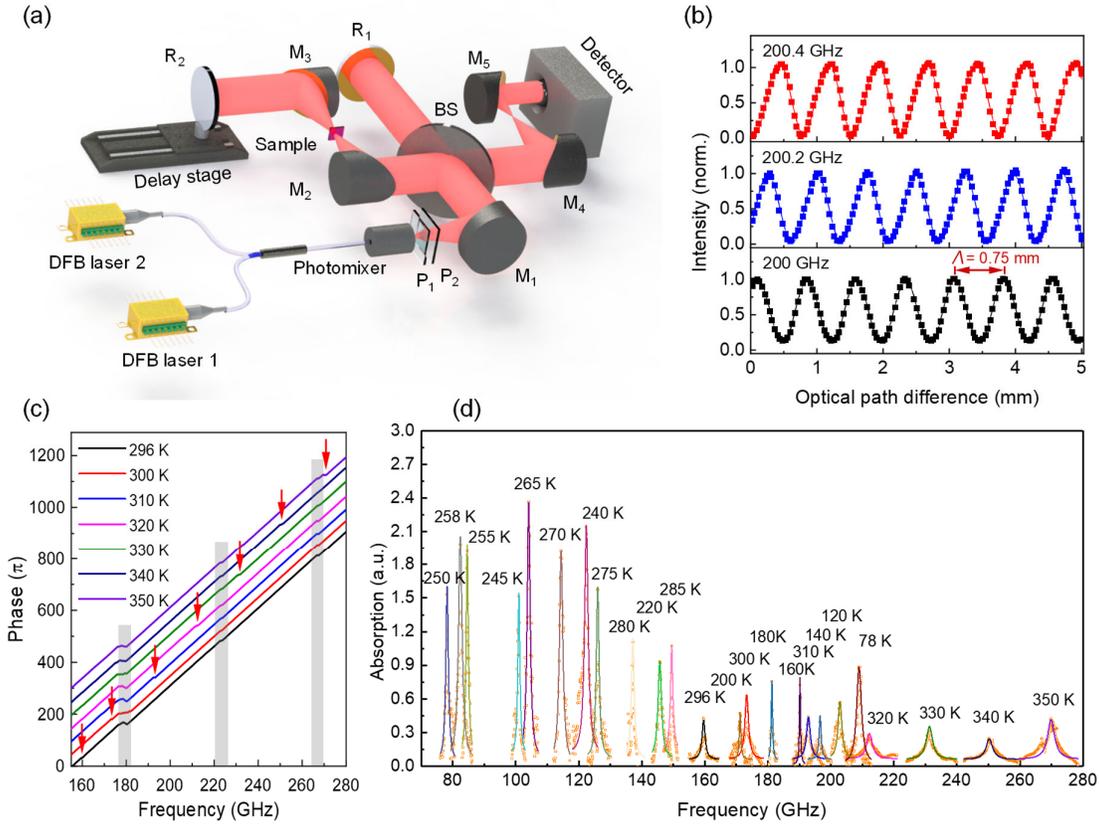

**Figure 2.** CW THz interferometer and phase-sensitive spin resonance measurements. a) An experimental schematic of the CW THz interferometer. A THz beam in red is generated at a photomixer by a heterodyne technique using two distributed feedback (DFB) lasers. Two polarizers ($P_{1,2}$) are used to tune the THz polarization. A highly-resistive silicon thin plate is used as a beam splitter (BS). The THz beams are guided by two plane gold mirrors ($R_{1,2}$) and four off-axis parabolic mirrors ($M_{1,2,3,4}$). $R_2$ is located on a highly accurate delay stage for scanning the optical path difference. The detector is a power meter and independent of the THz polarization. b) Typical interferometry data in the time domain at fixed THz frequencies of 200.0, 200.2, and 200.4 GHz are captured by a detector in a 5 mm delay stage scanning range at room temperature, where $\Lambda$ represents the period of the sinusoidal fits, e.g. $\Lambda$ = 0.75 mm at 200 GHz. c) The unwrapped phase at different temperatures. The red arrows indicate the spin resonances. Fabry-Pérot signals irrelevant to the spin resonance are shown by grey bars. The curves are vertically shifted with interval distances of $50\pi$ for clarity. d) Absorption derived from the phase. The Lorentzian fits are in solid lines.

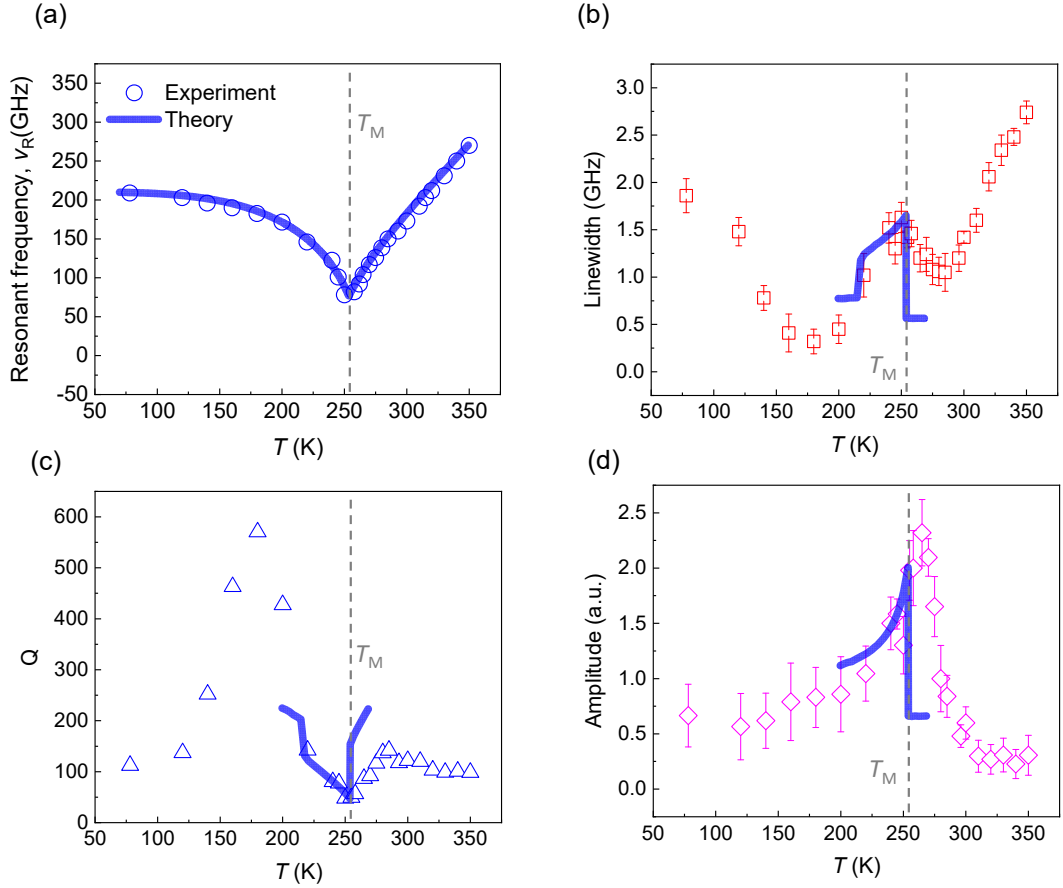

**Figure 3.** Antiferromagnetic resonance as a function of temperature. a) Temperature ($T$) dependence of the resonance frequency from 78 to 350 K. b) Temperature dependence of the linewidth (the full-width at half-maximum of Lorentzian fitting of the resonance absorption). c) Derived $Q$ factor from (a) and (b) with the maximum value of 570 at 180 K. d) The temperature-dependent resonance amplitude. The grey dash lines indicate the $T_M$ and the solid blue lines are theoretical fits by the spin-wave model.

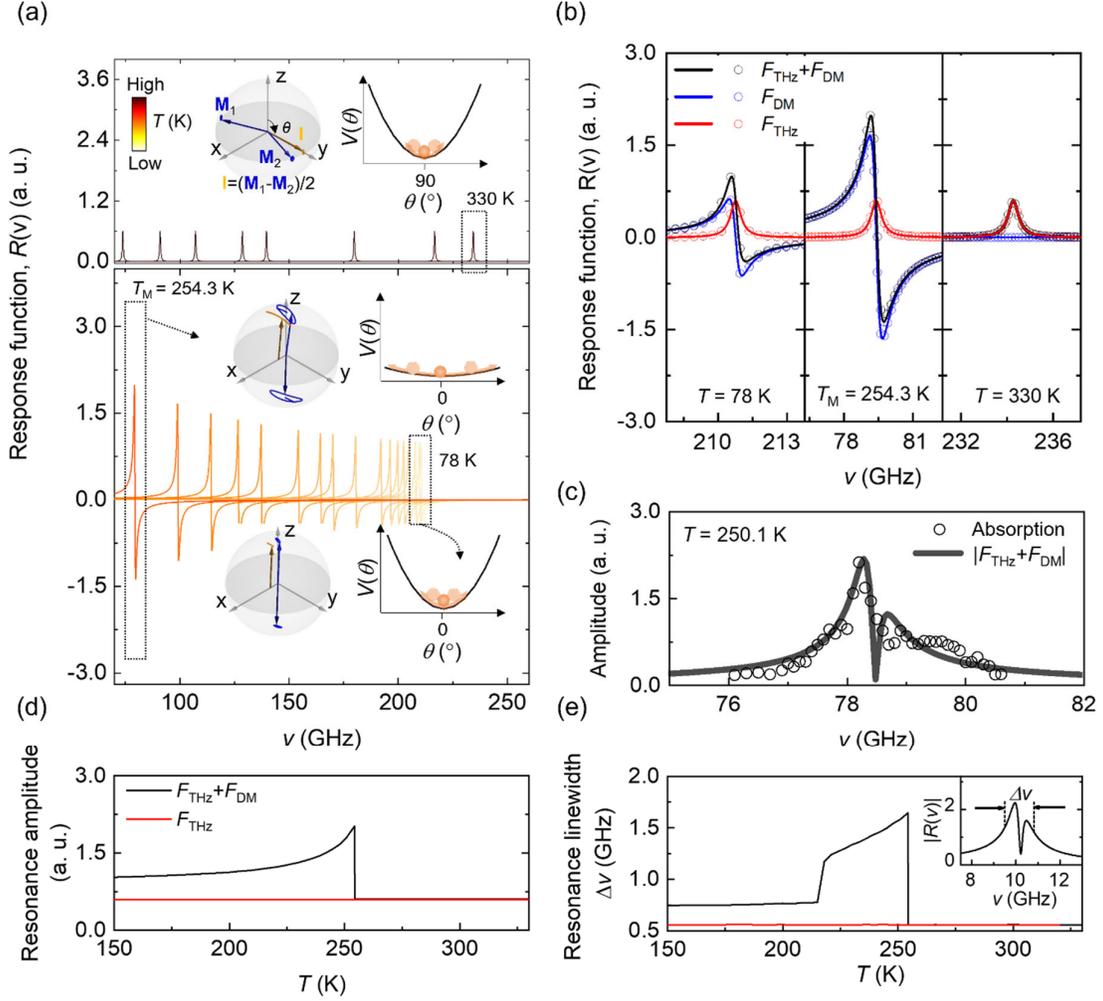

**Figure 4.** Simulated spin dynamics near the transition. a) The response function $R(v)$ induced by the DM interaction-induced driving force, $F_{DM}$ and THz magnetic field-induced driving force, $F_{THz}$. At $T < T_M$ ($T_M = 254.3$ K) and $T > T_M$, the resonance amplitude is low due to the dominant $F_{THz}$ and large potential curvatures (see insets). Near $T_M$, the resonance amplitude is high due to the dominant $F_{DM}$ and small potential curvature. Insets: the spin trajectories (left) of the sublattices $\mathbf{M}_1$ and $\mathbf{M}_2$ (blue arrows) and the Néel vector $\mathbf{l}$ (orange arrow), and magnetic potential (right) as a function of Néel angle, $\theta$ at 78, $T_M$, and 330 K. b) The spectrum analysis of $R(v)$ at 78, $T_M$, and 330 K. Red (Blue) line shows that $F_{THz}$ ($F_{DM}$) induces symmetric (anti-symmetric) spectrum. The symbols are from Eq. (2), and the lines are from the theoretical response function. c) The absorption data (black open circles) around $T_M$ and extraction of DM interaction from a fit (solid line) of the absorption data. d) The calculated resonance amplitude. e) Calculated resonance linewidth, $\Delta v$ of the full-width at half-maximum of the response function envelope (see inset), showing a peak near $T_M$.